\documentclass{PoS}
\usepackage{amsmath,wrapfig,cite}
\title{Recent development in parton shower multijet merging}

\ShortTitle{Recent development in parton shower multijet merging}

\author{\speaker{Johannes Bellm}\\
        IPPP, Department of Physics, Durham University\\
        E-mail: \email{johannes.bellm@durham.ac.uk}}

\abstract{Higher order calculations are necessary to predict and describe measurements in high energy collider physics. In recent years multiple approaches to combine multiple next-to-leading (NLO) order corrections with parton showers had been presented. We present on recent developments and future perspective. We highlight similarities and ambiguities in the procedure of achieving a  multijet merging at NLO.}

\FullConference{XXIV International Workshop on Deep-Inelastic Scattering and Related Subjects\\
		11-15 April, 2016\\
		DESY Hamburg, Germany}

\begin{document}

\section{Introduction}
\begin{wrapfigure}{r}{0.5\textwidth}
  \begin{center}
    \includegraphics[width=0.48\textwidth]{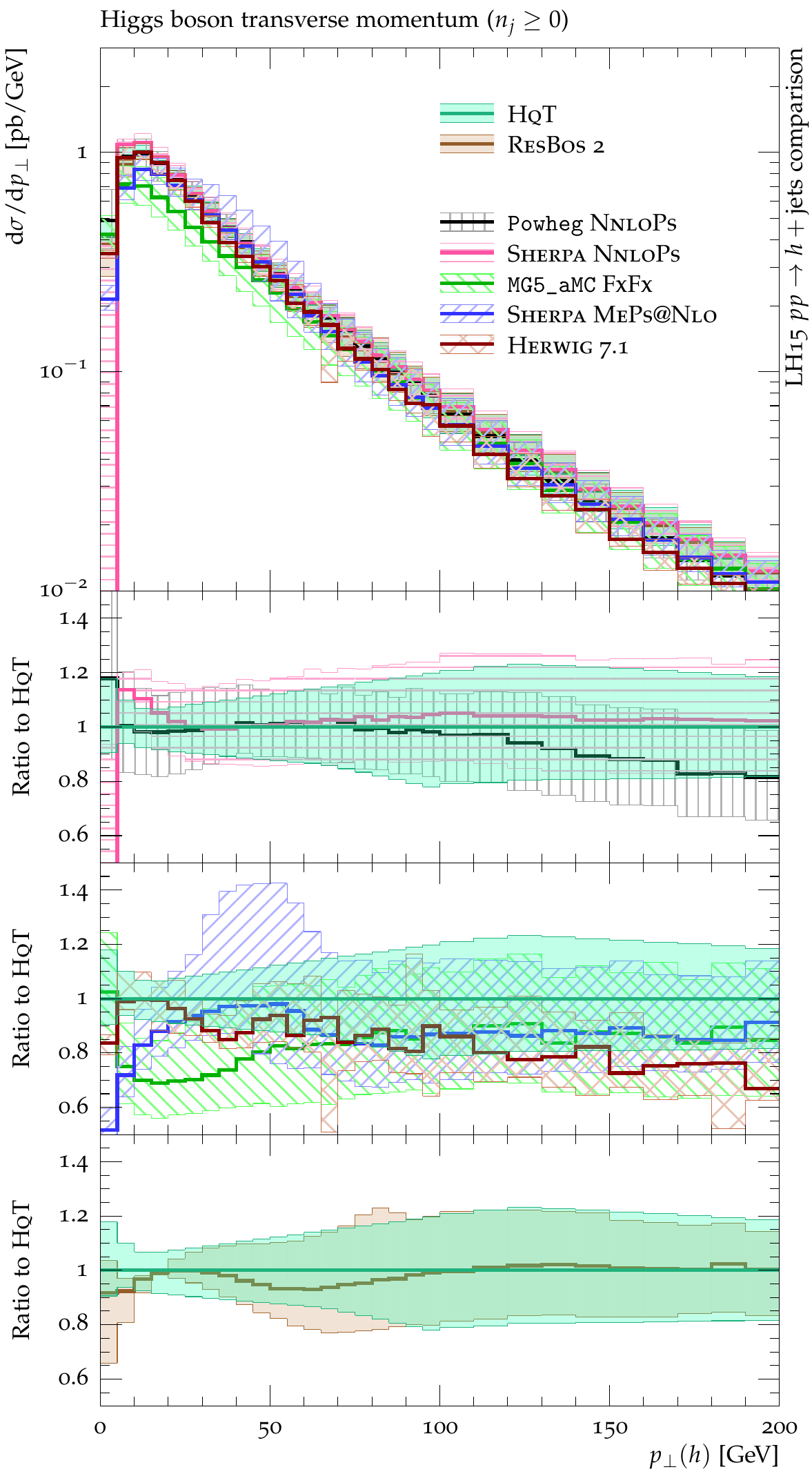}
  \end{center}
  \caption{The transverse momentum of a Higgs boson as presented in \cite{LesHouches15}. Multiple NLO merging schemes but also NNLO+PS  and analytic resummed results are compared and agree in general within the uncertainties. \label{fig:inclHiggsPt} }
\end{wrapfigure}
In the era of LHC the desire and necessity for higher precision predictions of exclusive observables increased. 
After the discovery of the Higgs boson \cite{HiggsAtlas,Higgs:CMS} the next steps are to precisely measure the properties, search for possible deviations to the standard model and assign uncertainties to the results \cite{LOShowerUncertainties,MEPSatNLOUncertanties}. 
Monte Carlo event generators \cite{Herwig,Herwig7,Pythia8.2,Sherpa} are  an essential tool in the comparison to the data taken 
at the LHC.
Relying on  leading-order (LO) cross section predictions, they evolve with parton shower approximations from a hard scale to the scales of hardonization. 
In order to describe hard and wide angle radiation with LO accuracy, first matrix element corrections \cite{Mike} and later  the method of merging multiple LO expectations\cite{CKKW,CKKW-L, MLM,CKKW2trunk,Lonnblad:2011xx} have been derived and implemented in event generators.
Parallel to the introduction of merging multiple LO contributions, the inclusion of NLO corrections to the cross section has been developed \cite{MCatNLO,POWHEG}.
Various implementations are available \cite{POWHEGBOX,aMCatNLO,Herwig7,Sherpa} and with the automation of one-loop calculations \cite{aMCatNLO,OpenLoops,GoSam,GoSam2,Actis:2016mpe} and infrastructures to combine them  with the event generation \cite{BLHA2} the techniques are widely used for studying physics at the LHC and other colliders. 
In the last years the development is driven by the availability of NLO corrections and the aim to combine more than one NLO correction with the parton shower approximation. In this talk we give an introduction on the techniques of this multijet NLO merging an overview on recent developments.

\section{Review on LO merging and NLO matching}
\label{sec:LOmergingandNLOmatching}
When it comes to improving the all-order but approximated description of a LO+parton shower (LO+PS) prediction with correct but fixed order matrix elements, algorithms need to be constructed to neither spoil the PS, nor the fixed order accuracy. PS algorithms iteratively  produce   emissions according to
\begin{equation}
PS_Q[d\sigma(Q)u(\phi_n)]=d\sigma(Q)\Delta(Q,\mu)u(\phi_n)+\int P(z,q)\Delta(Q,q)d\sigma(Q)PS_q\left[u(\phi_{n+1})\right]\;.
\end{equation}
In this condensed notation, the PS 'operator' $PS_Q[]$ acts on a multi particle state $u(\phi_n)$ by either not emitting an additional parton or an emission is produced according to $ P(z,q)\Delta(Q,q)dq$. Here $\Delta(Q,q)$ is the Sudakov form factor. 
In the collinear/soft limit $P(z,q)$ is constructed to reproduce  $P(z,q)\sigma_{n}(Q)\approx d\sigma_{n+1}(Q)$. 
While the fixed order contributions to higher multiplicities are divergent when additional emissions reach the collinear/soft limit and need to be regularized by cuts on the phase space, in PS the no-emission probability regularizes these regions. 
To include fixed order corrections to the approximation made in a PS, it was proposed to stack multiple fixed order contributions by reweighting the several contributions with the no-emission probability and factors to reproduce the scale dependent  factors ($\alpha_S$- and PDF-ratios) used in a PS algorithm \cite{CKKW,CKKW-L}.  

As a simplified general formula we use,
\begin{equation}
PS_Q[d\sigma(Q)_{\rm merged}u(\phi_n)]=\sum_i d\sigma_i(Q)\Delta(Q,\mu)u(\phi_i)+ d\sigma_N(Q)\Delta(Q,q)PS_q\left[u(\phi_{N})\right]\;.
\end{equation}
Here the individual expressions are reweighed with appropriate shower histories in order to render the expressions exclusive. The highest multiplicity undergoes the PS algorithm in order to produce higher multiplicities than are included as LO contributions.

More recently, and with the aim to include multiple NLO corrections, LO merging algorithms have been introduced to reproduce inclusive observables by not only  changing the emission probability (stacked higher multiplicities) but also the no-emission contribution\cite{ULOPS,HerwigMerging}. 
In this unitarized merging algorithms the no emission contribution of multiplicity $n$ is constructed by subtraction of the emission contribution of multiplicity $n+1$, so the replacement

\begin{equation}
 d\sigma_i(Q)\Delta(Q,\mu)u(\phi_i) \to d\sigma_i(Q)\Delta(Q,q_i)u(\phi_i) - \int_\mu^{q_i} d\sigma_{i+1}(Q)\Delta(Q,q_{i+1})u(\tilde{\phi}_i)\;.
\end{equation}

 is performed.

Another method to include, not only, matrix element corrected emissions but also achieving   NLO accuracy, is realized in NLO matching schemes. 
The most popular and widely used are the MC@NLO and POWHEG methods\footnote{Recently the KRKNLO method was introduced\cite{Jadach:2015mza}.}. 
In contrast to the merging procedures the corrections are included, not by replacing certain parts, but by adding the correction and subtracting the $\mathcal{O}(\alpha_S)$-expansion of the shower expressions\footnote{In the POWHEG method the shower is modified in a way, to produce the first emissions by $ME$ corrections.}. By these procedures the problem of double counting of the $\mathcal{O}(\alpha_S)$ contributions already approximated by the PS   is solved.

\section{From LO to NLO merging}
\label{sec:LOtoNLO}

The inclusion of multiple NLO corrections to the LO merging prescriptions require a closer insight in the LO merging schemes. 
Since in LO merging  $\mathcal{O}(\alpha_S)$  expressions are partly generated by the   LO contributions of the higher multiplicities, parts of the NLO corrections are already included.
The naive addition of NLO corrections would generate a similar double counting as addressed in the matching schemes. 
\begin{wrapfigure}{r}{0.5\textwidth}
  \begin{center}
    \includegraphics[width=0.48\textwidth]{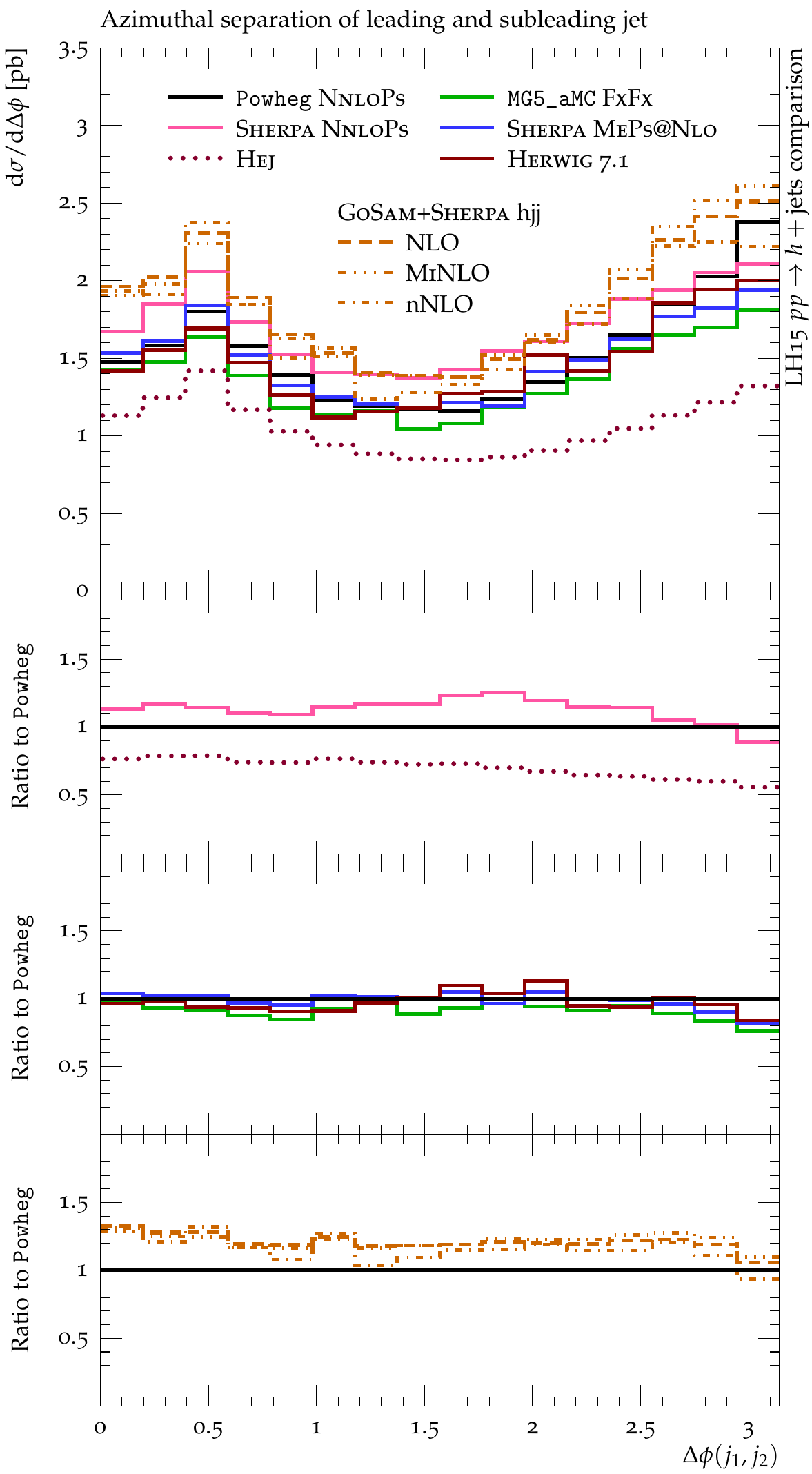}
  \end{center}
  \caption{Azimuthal difference of the two hardest jets in Higgs+jets events as presented in \cite{LesHouches15}. The NLO merged samples  describe the second emission with NLO accuracy and nicely agree for this observable.\label{fig:deltaphijj}}
\end{wrapfigure}
In order to solve the problem for multiple NLO corrections, all  $\mathcal{O}(\alpha_S)$ contributions generated by the LO merging need to be reviewed.

One of the schemes combining the LO merging ideas with fixed order NLO corrections is the MiNLO method \cite{MiNLO,MiNLOprime}. Here e.g. the Higgs transverse momentum is predicted with a fixed Order NLO calculation weighted with analytical Sudakov factors.
 In addition the MiNLO approach needs -- as the following algorithms -- to expand the Sudakov expression to $\mathcal{O}(\alpha_S)$ such as to restore NLO accuracy of the spectrum. 
 Since LO is modified by $\mathcal{O}(\alpha_S)$ expressions, the inclusion of NLO corrections require to subtract these. 
 
 Another proposal, on which the unitarized  NLO merging algorithms are based on is the LoopSim method\cite{LoopSim}. 
 Here multiple fixed order NLO calculations are combined by projecting multiplicity $n>1$ to a $n-1$-parton configuration. Since no PS and Sudakov suppression is used, the expansion is not required. 
 
In general the NLO merging algorithms have in common that the master formula for the additional  terms to achieve NLO accuracy are of similar form. Extraction of the $\mathcal{O}(\alpha_S)$ components for multiplicity $n$ and $n+1$ one gets,

\begin{align*}
d\sigma_n(Q)\Delta_{n}(Q,\mu)u(\phi_n)&\longrightarrow d\sigma_n(Q)\frac{\partial\Delta(Q,\mu)}{\partial \alpha_S} \alpha_S u(\phi_n)\\
d\sigma_{n+1}(Q)\Delta_{n+1}(Q,\mu)u(\phi_{n+1})&\longrightarrow d\sigma_{n+1}(Q)u(\phi_{n+1})
\end{align*}
where the first line produces a shower approximation similar to the MC@NLO procedure and  requires an additional expansion of intermediate Sudakov form factors as well as $\alpha_S$ and PDF ratios. 
The second line creates a subtraction for the real emission contribution of the NLO contribution. 

The different merging algorithms, as they are described below, alter in the way they implement these formulae. Here the difference is as in the choice of a shower algorithm of higher order as the expected accuracy. However, the treatment of merging scales, scale setting, cluster algorithms and the PS algorithms to which they are adapted impact the results. 
 
\section{Schemes }
\label{sec:schemes} 

\begin{wrapfigure}{r}{0.5\textwidth}
  \begin{center}
    \includegraphics[width=0.48\textwidth]{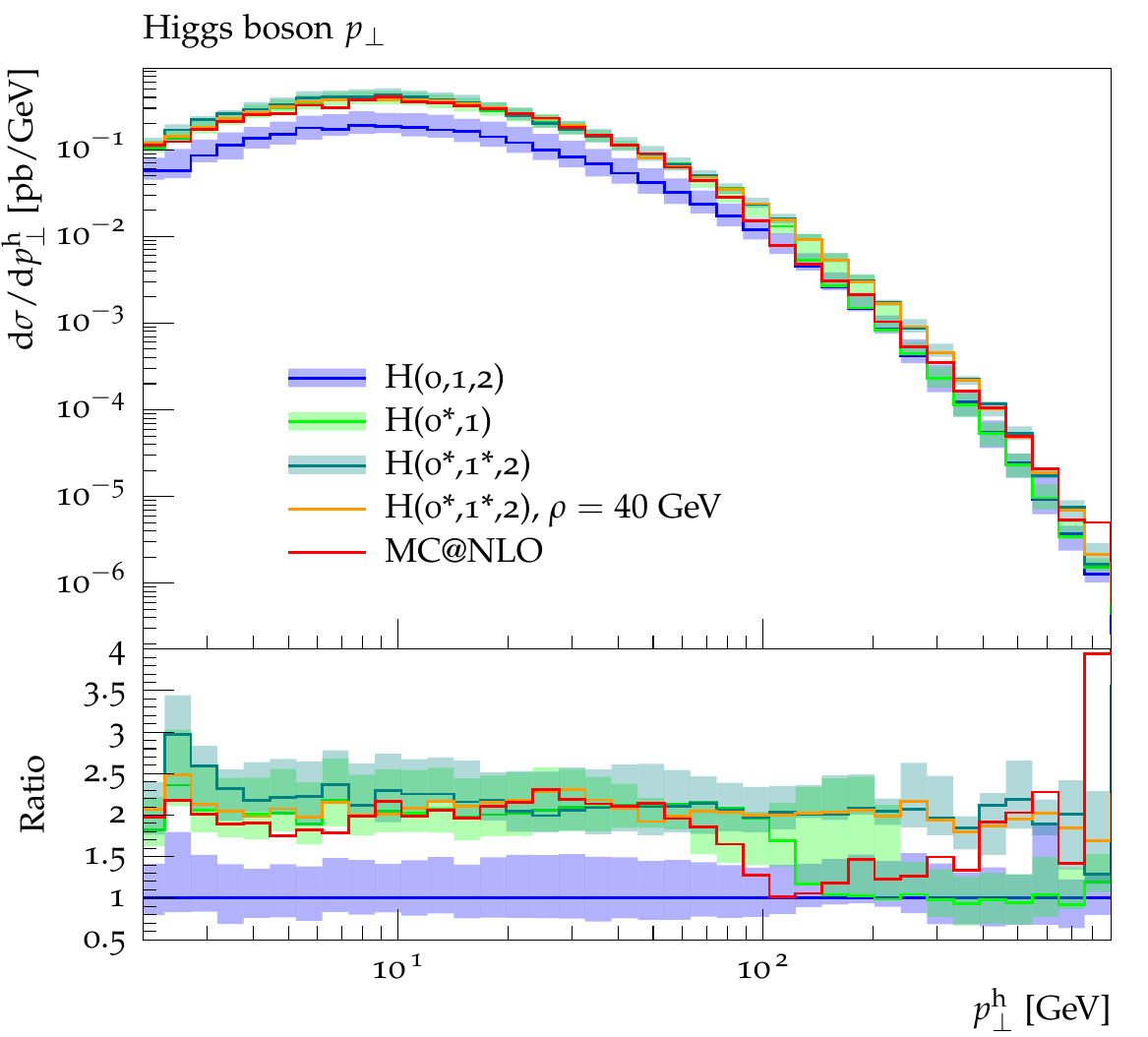}
  \end{center}
  \caption{The LO merged sample $H(0,1,2)$ with up to two jets at LO is compared with $MC@NLO$ and NLO merged samples with corrections to the production process and one additional emission.   Plot is taken from \cite{PhDBellm}.\label{fig:HiggsPt}}
\end{wrapfigure}

In this section we describe the main features of  several schemes and point out some key ingredients. 
NNLO+PS combinations typically inherit same merging features, see contribution of S. Alioli\footnote{ 
The list of schemes is not complete, e.g. the GENEVA \cite{GENEVA} approach.}.

 NL$^3$:  
The first complete implementation of a NLO merging scheme is called $NL^3$\cite{NL3}. 
It was applicable for $e^+e^-$ collisions and was later extended to hadron collisions \cite{UNLOPS}. An algorithm is provided to calculate the expansion of the Sudakov form factor to $\mathcal{O}(\alpha_S)$. The NLO corrections are included  as overall factors $k_i$ for each corrected multiplicity. $k_i$ are calculated with integrated real emission and subtraction expressions below a cut $y_{cut}$.

MEPS@NLO:
A more differential approach of the NLO corrections was presented in the approach to stack multiple $MC@NLO$ contributions in \cite{MEPSatNLOLEP,MEPSatNLO}. This implementation in the Sherpa framework creates the expansion of the Sudakov exponent by reweighting the showering algorithm and makes use of shower approximations for real subtraction and truncated showering to fill the phase space not contained in the matrix element region.

FxFx:
As MEPS@NLO, the FxFx \cite{FxFx } scheme makes use of the more differential approach of stacking multiple MC@NLO combinations which are in this case combined with smooth separation factors  instead of a hard merging scale. Here the calculation of the Sudakov suppression is  a mixture of  analytic NLL-CKKW \cite{CKKW} like and a MLM-like\cite{MLM} rejection. The usual $\alpha_S$-ratios, as they appear in CKKW-like procedures is here approximated by using a scale setting similar to MiNLO\cite{MiNLO}.

UNLOPS:
While the previous schemes are extensions of  usual LO merging schemes, the UNLOPS method  unitarises the LO merging by subtracting the same weights as they appear in the emission expressions\cite{ULOPS,HerwigMerging,UNLOPS}. LO merging is able to change the inclusive cross section at the $\mathcal{O}(\alpha_S)$-level, which is in this scheme cured by construction. Another key ingredients is the treatment of the PDF ratios, which appear in the LO version of the CKKL-L merging.

Merging in Herwig:
As in UNLOPS a unitarisation of the LO merging is used in the Herwig merging, based on \cite{HerwigMerging} and \cite{PhDBellm},  in order to restore the LO cross section.
Below the merging scale MC@NLO-like subtracted and differential real emission contributions fill the phase space. 
The Sudakov suppression and expansion can be calculated numerically and with trial emissions for cross checks.
In addition the PS algorithm is modified to accept emissions in regions where wide angle emission become soft/collinear to other partons. 

\section{Example Results}
\label{sec:results}

 The schemes are implemented in event generators and studies have been made to test the algorithms and the difference in between. Fig. \ref{fig:HiggsPt} shows the Higgs bosons transverse momentum in a LHC environment. The underlying simulation is the Herwig merging. 
 As a comparison the matched NLO+PS in the MC@NLO scheme as it is implemented in Herwig7 is shown. 
 The NLO corrections in the MC@NLO scheme produce a strong increase in the shower phase space of the first emission. The NLO merged sample with a single one-loop correction to the production process indicated with $H(0^*,1)$ behaves like the MC@NLO contribution. When a NLO correction to the contribution with an additional emission is merged ($H(0^*,1^*,2)$) a nearly constant ratio compared to the LO merged prediction ($H(0,1,2)$)  is received for this process in this scheme. The difference between MC@NLO or  $H(0^*,1)$ to $H(0^*,1^*,2)$ is driven by the large correction to the H+1jet cross section. 
 
 An exhaustive study to test the differences in NLO merging between the schemes for Higgs production have been presented in \cite{LesHouches15}. Figs. \ref{fig:inclHiggsPt} and \ref{fig:deltaphijj} show again the transverse momentum of the Higgs boson an the azimuthal difference of the two hardest jets. 
 Within the error bands, which have been produced by varying the renormalisation and factorisation scale of the hard process, the predictions mostly agree. Also the comparison to analytic resummation and NNLO+PS matching was studied. 
 In \cite{LesHouches15} other distributions are discussed and larger differences are observed when more exclusive observables are reviewed.

\section{Summary and Outlook}
\label{sec:summary}
An overview on the different schemes to merge multiple NLO corrections with parton shower predictions was given. Therefore a general overview on the path to correct parton showers with fixed order calculations was explained. In recent years the understanding of subject was improved. It is important to perform comparisons of the different schemes, which should only differ in subleading expressions, in order to  find differences and produce reliable predictions.

\section*{Acknowledgements}
The author thanks S. Gieseke and S. Pl\"atzer for collaboration on  multijet merging in Herwig and  F. Krauss, L. L\"onnblad, and S. Prestel  for fruitful discussions. The invitation of the DIS16 organizers is gratefully acknowledged. I thank S. Pl\"atzer for carefully reading the manuscript. 
This work was supported  by the European Union as part of the FP7
Marie Curie Initial Training Network MCnetITN (PITN-GA-2012-315877). 

\bibliography{PoS-DIS16-Bellm}

\end{document}